\documentclass[showpacs,11pt,A4paper]{revtex4}
\usepackage{graphicx}

\begin{document}

\title{Electron-lattice interaction and structural stability of the oxy-borate $Co_3O_2BO_3$}
\author{}
\author{M. Matos}

\affiliation{Departamento de F\'\i sica, PUC-Rio, G\'avea, CEP 22453-970,
Caixa Postal 38071, Rio de Janeiro, RJ, Brazil}

\begin{abstract}

A theoretical study is carried out in the homometallic mixed valent ludwigite $Co_3O_2BO_3$ using a modified tight binding methodology. The study focuses on the electronic properties of bulk, 1D and molecular units to describe differences between $Co_3O_2BO_3$ and another homometallic ludwigite, $Fe_3O_2BO_3$. The latter is known to present a structural instability which has not been found in $Co_3O_2BO_3$. Our results show that bulk band structures present no significant differences. Differences are found in the calculation of 1D stripes formed by $3^+ 2^+ 3^+$ triads, owed to different $3d$ occupancy of $Fe$ and $Co$ cations. Conditions for 1D semiconducting transport observed in $Fe_3O_2BO_3$ are not present in $Co_3O_2BO_3$. More important differences were found to be related to local octahedral geometry. Larger distortions in $Co_3O_2BO_3$ lead to higher $2^+ \rightarrow\ 3^+$ hopping  barriers in the triads and consequent localization of charge. In $Fe_3O_2BO_3$, site equivalence provides easy paths for inter-cation hopping. It is then suggested that local geometry of the cation sites is the main cause of differences in these two compounds; dimerization in the triads, which characterizes the structural change in $Fe_3O_2BO_3$, could be structurally hindered in $Co_3O_2BO_3$. An analogy is made with two mixed-valent {\it warwickites}, $Fe_2OBO_3$ and $Mn_2OBO_3$, which show the same structural stability/site equivalence relationship.

\end{abstract}

\maketitle


\section{Introduction}

Ludwigite is an oxyborate with formula unit $M'_2MO_2BO_3$ in which divalent and trivalent metals $M'$ and $M$ are found inside $4$ non equivalent edge-sharing oxygen octahedra, forming zig-zag planes mutually linked through $BO_3$ trigonal groups\cite{takeuchi}. $Fe$ and $Co$ homometallic ludwigites are mixed valent compounds and were synthesized for the first time about two decades ago\cite{swinnea,norres-co}. $Fe_3O_2BO_3$ turned out to be the most challenging compound of the ludwigite family\cite{att92,wigner-glass,larrea-local,douvalis02,uff04-magn,uff05-transp}. In the last few years increasing interest has been devoted to this material, motivated by its complex electronic, magnetic and transport properties and to a broad ($150$ to $300K$) orthorhombic-orthorhombic structural change\cite{struc-tr,bordetsuard}. Its main effect involves a distortion in a triad of $Fe^{3+}$-$Fe^{2+}$-$Fe^{3+}$ cations with small ($2.6-2.9 \AA$) inter-atomic distances. The pilling up of $Fe$ triads along the short orthorhombic $c$ axis has an almost independent 1D character in many respects and has been referred to as a 3LL (three-leg-ladder) sub-structure, immersed in the corrugated 2D planes\cite{latge}. 

The structural transition in $Fe_3O_2BO_3$ has been theoretically investigated\cite{latge} and it has been suggested that its origin is associated to excitonic instabilities in the 3LL substruture. In another approach, based on orbital interactions, it was found that $3d$-$2p$ hybridization leads to small hopping barriers between $Fe^{2+}$ and $Fe^{3+}$, favouring charge ordering at low temperatures\cite{mm-ludwfe}. To improve comprehension on the mechanisms involved, attention has been turned more recently to the other homo-metallic ludwigite, $Co_3O_2BO_3$\cite{ludwco1,ludwco2}. Despite the structural similarity between the two compounds, no conformational change was found in the latter. $X$-ray diffraction measurements, performed at low ($105 K$) and high ($293 K$) temperatures\cite{ludwco1}, showed the same orthorhombic structure, in the {\it Pbam} space group, found earlier by Norrestam et al.\cite{norres-co}. 

Magnetic properties present similarities and differences in the two materials. Both exhibit a low temperature magnetic order. $Co_3O_2BO_3$ becomes a weak ferromagnet or ferrimagnet below $\sim 45 K$\cite{ludwco1,ludwco2} and $Fe_3O_2BO_3$ shows antiferromagnetic (AF) order below $50K$\cite{uff04-magn,bordetsuard}. The $Fe$ mixed-valent material has another AF phase, from $70K$ to $112K$, in which only part of the atoms are magnetically aligned\cite{uff04-magn,douvalis02}. Between $50$ and $70K$ the passage from partial to total alignment of $Fe$ spins (attained at $50K$) is set in through a weak ferro-\cite{uff04-magn} or ferri-magnetic\cite{douvalis02} order. The differences in magnetic and structural behavior of the two ludwigites would suggest that magnetism is involved in the structural change observed in $Fe_3O_2BO_3$. Nevertheless, the fact that in the iron material structural and magnetic transition temperatures are relatively far appart has been considered as an indication that the mechanism of atomic rearrangements is not directly and uniquely related to magnetism\cite{ludwco1}.

Electrical transport is more closely related to the structural transition in $Fe_3O_2BO_3$. In the low temperature phase, conductivity was found by different authors to be thermally activated, with activation energy of $\sim 0.2 eV$\cite{struc-tr,uff05-transp,douvalis02}. At high temperature, the conductivity regime changes, with a steep decrease in the logarithmic derivative of the resistivity as a function of temperature\cite{uff05-transp}. For $Co_3O_2BO_3$, two mechanisms of electrical transport have been proposed\cite{ludwco1,ludwco2}.  

Charge distribution in $Fe_3O_2BO_3$ has been intensively studied through M\"ossbauer spectroscopy and a clear pattern was found in a temperature range as wide as $4K<T<600K$\cite{swinnea,larrea-local,uff04-magn,douvalis02}. Below $\sim 50K$ complete ordering is found with all $Fe$ atoms having valence $+2$ or $+3$, in the proportion $2:1$ as expected from stoichiometry. The trivalent metal sits preferentially at the borders of the triads. From this low temperature and up to $\sim 300K$, intermediate valence $+2.5$ is observed in the triad, indicating the presence of one $Fe^{2.5+}-Fe^{2.5+}$ pair and one trivalent $Fe^{3+}$ atom per triad. This result indicates rapid hopping in one $Fe$ dimer per triad. Above $300K$ the $Fe^{2.5+}$ disappears from the M\"ossbauer spectra and again only divalent and trivalent $Fe$ are found. Simultaneously, delocalization along $c$ involving the extra $Fe^{2+}$ electron was detected\cite{larrea-local,uff04-magn}. The onset of 1D charge transport in the triad, along the $c$ axis was previously suggested by Swinnea et al.\cite{swinnea}. Douvalis and colaborators\cite{douvalis02} have observed a valence of $+2.75$ in the triad just before the disappearance of intermediate valences (at $\sim 270K$). This indicates that near the transition temperature, the extra $Fe^{2+}$ electron gets more delocalized in the triad, so that one has one pair with $+2.75$ and one atom with $+2.5$. This delocalization inside the triad precedes 1D transport along the 3LL, in the non-dimerized orthorhombic structure. 

An analogous behavior is found in mixed-valent warwickites, $Fe_2OBO_3$ and $Mn_2OBO_3$. The former undergoes a broad structural transition at $317 K$, associated with a change in the conductivity regime and charge rearrangement in the low temperature phase\cite{nature,att99-warwfe}. $Mn_2OBO_3$, however, presents no structural transition. In this compound, strong Jahn-Teller distortions of $Mn^{3+}$ lead to a charge ordering state independent of temperature\cite{norres-mn,nature,goff}. 

In the present paper, the extended H\"uckel tight binding (EHTB) method is used to investigate the electronic structure of $Co_3O_2BO_3$, using the high spin band $\it hsf$ approach. Comparison is made with the electronic structure of the high temperature phase of $Fe_3O_2BO_3$, whose crystal structure is similar to that of the cobalt material. The calculations on $Fe_3O_2BO_3$ add to former calculations in the system\cite{mm-ludwfe,mm-souza}. Norrestam et al.\cite{norres-co} gave a previous account of EHTB calculations on  $Co_3O_2BO_3$. The $\it hsf$ procedure prevents unphysical excesses of electron population in low lying $3d$ levels of the metal cation, allowing a correct account of basic structure-properties relationship. It is appropriate to narrow $d$ band systems and showed to be particularly suitable to describe $Fe_3O_2BO_3$\cite{mm-ludwfe} and other oxy-borates\cite{mm-warwmn,mm-warwfe}. 

\section{Theory}

The extended H\"uckel theory\cite{eHT,solidsurfaces} is widely known in the literature; here a brief summary of the relevant aspects of the theory, to the present study, is given. For the extended H\"uckel hamiltonian, diagonal terms $H_{ii}$ and overlap integrals $S_{ij}$ between atomic orbitals of the Slater type are needed. $S_{ij}$ overlaps take into account the particular geometry of the system. The parameters used in the present study are chosen from ref.\cite{tables} and given as follows: (i) for $O$, $H_{2s,2s}$=$-32.3 eV$, $H_{2p,2p}$=$-14.8 eV$, $\zeta _{2s}$ = $\zeta _{2p}$=$2.275$; (ii) for $B$, $H_{2s,2s}$=$-15.2 eV$, $H_{2p,2p}$=$-8.5 eV$, $\zeta _{2s}$=$1.3$, $\zeta _{2p}$= $1.2$; (iii) for $Mn$, $H_{4s,4s}$=$-9.75 eV$, with $\zeta _{4s}$=$1.8$, $H_{4p,4p}$=$-5.89 eV$ with $\zeta _{4p}$=$1.8$, $H_{3d,3d}$=$-11.67 eV$, $\zeta_1$ = $5.15$, $c_1$=$0.5113$, $\zeta_2$=$1.90$, $c_2$=$0.6659$; (iv) for $Fe$, $H_{4s,4s}$=$-9.10 eV$, with $\zeta _{4s}$=$1.9$, $H_{4p,4p}$=$-5.32 eV$ with $\zeta _{4p}$=$1.9$, $H_{3d,3d}$=$-12.6 eV$, $\zeta_1$ = $5.35$, $c_1$=$0.5505$, $\zeta_2$=$2.00$, $c_2$=$0.6260$; (v) for $Co$, $H_{4s,4s}$=$-9.21 eV$, with $\zeta _{4s}$=$2.00$, $H_{4p,4p}$=$-5.29 eV$ with $\zeta _{4p}$=$2.00$, $H_{3d,3d}$=$-13.18 eV$, $\zeta_1$ = $5.55$, $c_1$=$0.5679$, $\zeta_2$=$2.10$, $c_2$=$0.6059$. Off-diagonal terms of the hamiltonian matrix, $H_{ij}$, are determined from weighted averages of $H_{ii}$ and $H_{jj}$ taking also into account $S_{ij}$\cite{mm-theochem}. Density of states are obtained from a mesh of $192$ uniformly spread reciprocal lattice $k$ points, whose set was found to be well suited to the oxyborate systems\cite{mm-souza}. The Fermi level is defined as the highest occupied crystalline orbital. The high spin band filling ({\it hsf}) scheme has been discussed elsewhere\cite{mm-ludwfe}. An explanatory picture is that of minority spin electrons moving in a background of majority spin electrons. Mulliken population is used to calculate atomic charges.

All calculations were performed with bind\cite{bind} program and the graphics were drawn with viewkel\cite{viewkel}, distributed as part of the YAeHMOP package.

\section{Crystal structure}

Fig.1 shows a polihedral drawing of the ludwigite structure seen along the $c$-axis. Sites $1$,$2$ and $3$ are preferentially occupied by divalent metals while metal site $4$ is usually trivalent.  Two-fold $1$ and $2$ plus four-fold $3$ and $4$ provide eight di- and four tri-valent sites, thus satisfying the unit cell stechiometry $(M'^{2+}_2M^{3+}B^{3+}O_5^{2-})_4$. The $4$-$2$-$4$ triad is the basic unit that forms the 3LL 1D substructure, running paralel to the $c$-axis, mentioned above. Structure refinements of $Co_3O_2BO_3$ and of the $294K$ phase of $Fe_3O_2BO_3$ were made in the {\it Pbam} space group n. $55$ \cite{struc-tr,ludwco1,ludwco2}. Some structural parameters are given in Table 1, which contains an estimate of octahedral distortions $\Delta$, given by the average of the difference $[(d_{M-O_i}-d_{av})/d_{av}]^2$, $i=1,..,6$, for each $MO_6$ group\cite{shannon}.  

\begin{figure}[tbp]
\includegraphics[scale=0.7]{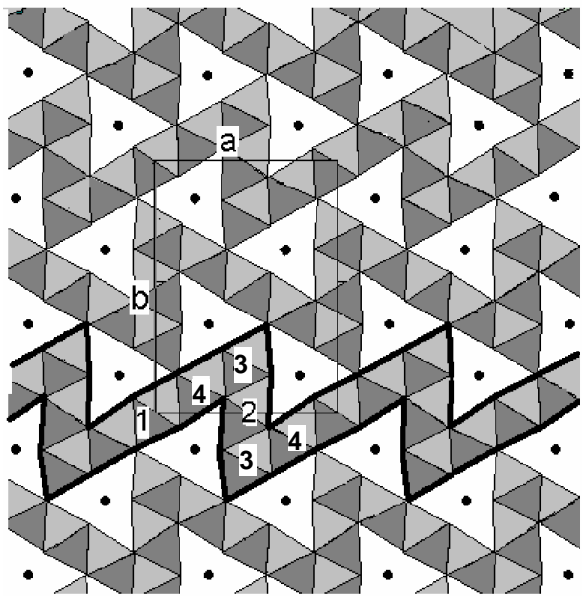}
\caption{The crystal structure of the ludwigite, seen along the hexagonal c-axis; $Fe$-triad: $4-2-4$.}
\label{fig1}
\end{figure}

\vskip.5in

\begin{tabular}{|c|c|c|c|c|c|}
\hline
\multicolumn{3}{|c|}{{\bf Table 1} - Structure data of homometallic ludwigites }\\ \hline
\multicolumn{3}{|c|}{Lattice parameters $a$,$b$,$c$, metal-oxygen average distances, }\\
\multicolumn{3}{|c|}{metal-metal distances (in \AA) and octahedral distortions $\Delta$}\\ \hline
& $Co_3O_2BO_3$    & $Fe_3O_2BO_3$  \\ \hline
$a$ & $9.30$&  $9.46$ \\ \hline
$b$ & $11.96$& $12.31$ \\ \hline
$c$ & $2.97$& $3.08$ \\ \hline
$d(M1-O)_{av}$ & $2.10$ & $2.15$  \\
$10^3\Delta$ & $1.04$ & $1.27$\\ \hline
$d(M2-O)_{av}$ & $2.07$ & $2.09$\\
$10^3\Delta$ & $0.81$ & $0.11$\\ \hline
$d(M3-O)_{av}$ & $2.09$ & $2.15$ \\
$10^3\Delta$ & $1.01$ & $1.92$\\ \hline
$d(M4-O)_{av}$ & $1.93$ & $2.06$ \\
$10^3\Delta$ & $0.15$ & $0.30$\\ \hline
$d(M1-M4)$ & $3.01$ & $3.10$\\ \hline
$d(M2-M3)$ & $3.06$ & $3.18$\\ \hline
$d(M2-M4)$ & $2.75$ & $2.79$\\ \hline
$d(M3-M4)$ & $3.09$ & $3.19$\\ \hline

\multicolumn{3}{|c|}{Data for $Fe_3O_2BO_3$ $\rightarrow$ $\Delta$: this work; other: ref. \cite{mm-souza}}\\ \hline
\end{tabular}
\vskip.5in

\section{Results}

The bulk (3D) and 3LL(1D) EHTB band structures of $Co_3O_2BO_3$ and $Fe_3O_2BO_3$ are shown in Fig.2 in the metal $3d$ energy range; band structure properties are given in Table 2. Oxygen $2p$ bands (not shown) appear at $\sim 0.9 eV$ ($Co$ ludwigite) and $\sim 1.5 eV$ ($Fe$ ludwigite) below the metal $d$ bands. The two separate groups of bands, seen in Fig.2, come from the cubic $d$-splitting of the oxygen octahedral field and will be denoted $t_{2g}$ and $e_g$.  Two narrower $t_{2g}$ bands are seen to be detached from the main group. They are characteristic of the ludwigite structure and come from strong metal-metal interactions in the triad, thus named $\sigma$ and $\sigma *$\cite{mmrh, mm-ludwfe}. The 3LL(1D) $0.22 eV$ semiconducting gap which appears in $Fe_3O_2BO_3$ (see Fig.2, fourth panel) has been associated to the electronic barrier responsible for the thermally activated behavior experimentally observed in the iron ludwigite\cite{mm-ludwfe}. The calculated gap is in good agreement with the experimental value, estimated to be $\sim 0.2 eV$\cite{mm-ludwfe,douvalis02,uff05-transp}. Its existence is also confirmed by M\"ossbauer data which indicates the onset of 1D transport through the 1D stripe\cite{larrea-local}. 

$Co^{2+}$ ($3d^{5+2}$) and $Co^{3+}$ ($3d^{5+1}$) give $8\times 2 + 4\times 1 = 20$ spin down electrons per unit cell, leading to $20/36$ fractional occupancy of $t_{2g}$ bands, larger than the $8/36$ fraction of $Fe_3O_2BO_3$. Therefore, due to electron counting, the Fermi level of the cobalt ludwigite is raised; a significant effect of electron counting is the disapearance of the 1D gap of $Co_3O_2BO_3$. This constitutes an important difference between both materials, from which a different conductivity behavior could be expected. This point will be discussed below.

\begin{figure}[tbp]
\includegraphics[scale=0.9]{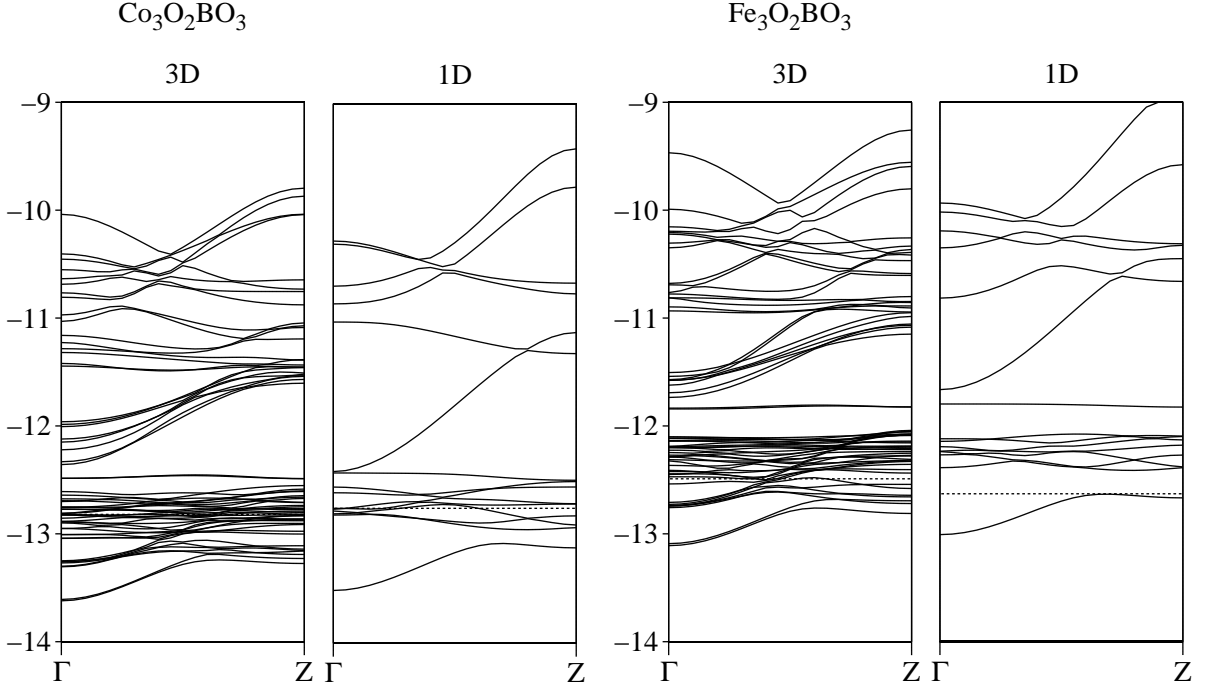}
\caption{The bulk and 1D (3LL) $3d$ bands of $Co_3O_2BO_3$ and $Fe_3O_2BO_3$, showing the $\sigma$ and $\sigma*$  bands. Note the 1D gap above the Fermi level (doted line) in the $Fe$ compound. Band structure of $Fe_3O_2BO_3$: see also refs.\cite{mm-ludwfe,mm-souza}.}
\label{fig2}
\end{figure}

\vskip.5in
\begin{tabular}{|c|c|c|} \hline
\multicolumn{3}{|c|}{{\bf Table 2} - Calculated electronic structure properties. Energies in eV.}\\ \hline
& $Co_3O_2BO_3$ & $Fe_3O_2BO_3$  \\ \hline
$E_F$ (1D) & $-12.75$&  $-12.63^*$ \\ \hline
$E_F$ (3D) & $-12.82$ & $-12.49$ \\ \hline
$\delta_c$ (1D) & $0.01$ & $0.14$  \\ \hline
$\delta_c$ (3D)  & $0.09$ & $0.07$ \\ \hline
1D gap& $0$ & $0.22$\\ \hline
\multicolumn{3}{|c|}{$E_F$: the {\it hsf} Fermi level; $\delta_c$: the $t_{2g}$-$e_g$ cubic splitting}\\ 
\multicolumn{3}{|c|}{* : top of $\sigma$ band}\\ \hline
\end{tabular}
\vskip.5in

In Table 3 are shown the calculated atomic charges. For consistency, one compares charge distribution in the crystal (3D) and the 3LL (1D) sub-system. 

\vskip.5in

\begin{tabular}{|c|c|c|c|c|c|}
\hline
\multicolumn{3}{|c|}{{\bf Table 3} - Metal charge distribution of $Co_3O_2BO_3$ and $Fe_3O_2BO_3$}\\ \hline
& $Co_3O_2BO_3^{(a)}$& $Fe_3O_2BO_3^{(b)}$ \\ \hline
\multicolumn{3}{|c|}{1D}\\ \hline
$q(M1)$ & $-$&  $-$ \\ \hline
$q(M2)$ & $0.777$& $1.255$ \\ \hline
$q(M3)$ & $-$& $-$ \\ \hline
$q(M4)$ & $1.529$ & $1.473$  \\ \hline
\multicolumn{3}{|c|}{3D}\\ \hline
$q(M1)$ & $0.848$&  $1.004$ \\ \hline
$q(M2)$ & $0.940$& $1.234$  \\ \hline
$q(M3)$ & $0.853$& $1.198$ \\ \hline
$q(M4)$ & $1.502$ & $1.235$  \\ \hline
\multicolumn{3}{|c|}{(a): this work; (b) ref. \cite{mm-ludwfe}}\\ \hline

\end{tabular}
\vskip.5in

Charge localization is clearly observed in $Co_3O_2BO_3$ in contrast with $Fe_3O_2BO_3$, where no significant distinction is obtained in $q(Fe)$ at different sites. In the former, calculated values of $q(Co1)$, $q(Co2)$ and $q(Co3)$ ($\sim 0.8-0.9$) and that of $q(Co4)$ ($\sim 1.5$), lead one to associate the former three to divalent and the latter to trivalent cations. Empirical estimates of cobalt oxidation state in $Co_3O_2BO_3$, using valence bond sums ({\it vbs}), provide for sites $1$, $2$ and $3$, valences of $1.91$, $2.06$ and $1.98$, respectively, consistent with $2+$ cations and $2.73$ for site $4$, which can be associated to a $3+$ oxidation state\cite{ludwco1}. In the 1D sub-systems, the same delocalization/localization behavior is observed in the $Co$/$Fe$ materials although, as expected, charges come out to be slightly more localized in both cases, due to lower dimensionality. Charge distribution is well correlated to the sizes of $O$ octahedra, given in Table 2. While in $Co_3O_2BO_3$, $2+$ octahedra are noticeably larger than $3+$, in  $Fe_3O_2BO_3$, the $Fe$ triad sites, besides being smaller than those outside, are very similar, consistent with intermediate valence. As expected, the quantum mechanical calculated charges differ in absolute values from empirical estimates but the same trend is found in the charge distribution. In the next section, other effects associated to octahedral subunits are investigated.       

\subsection{Monomers and dimers}

Basic insight of electronic processes taking place in the physical system is obtained by examining structural sub-units such as {\it monomers} $CoO_6$, $FeO_6$ and {\it dimers} $Co_2O_{10}$, $Fe_2O_{10}$. It has already been shown that the density of states of oxy-borates are well reproduced by the electronic levels of the respective sub-units\cite{mm-theochem,mm-warwfe,mm-warwmn,mmrh}. This is related to the ionic character of the system. As the hopping barriers are small, $\sim 0.1 eV$, we compare EHTB and more precise embedded cluster DFT calculations in the study of isolated subunits.


One considers first $M^{2+}O_6^{12-}$ and $M^{3+}O_6^{12-}$ isolated monomers ($M$=$Fe$, $Co$), carved out from the material's crystalline structure. In the high spin state of $Co^{2+/3+}$ ($3d^{5+2/5+1}$) there are $2/1$ minority spins so that $2/1$ molecular levels of the $t_{2g}$ group are doubly occupied. For $Fe^{2+/3+}$ ($3d^{5+1/5}$) one has $1/0$ minority spins in $1/0$ $t_{2g}$ levels.  In Fig.3, the calculated $t_{2g}$ molecular structure of the monomers is shown for the triad sites $2$ and $4$ of $Co_3O_2BO_3$ and $Fe_3O_2BO_3$. This is the energy range where comparison between sites is meaningful for the analysis. The occupied molecular orbital ($HOMO^{\downarrow}$) is indicated for each case ($Fe4$ is presented as a trivalent cation). Effects of geometrical distortions of the different oxygen octahedra are clearly seen to produce distinct degeneracy breaking of the $3$-fold $t_{2g}$ levels. 

From Fig.3, two elementary paths for intersite electron transfer between adjacent monomers in the cobalt $4$-$2$-$4$ triad can be indentified. One is from the $HOMO^{\downarrow}$ of $Co2$ to the $LUMO^{\downarrow}$ of $Co4$, namely $2\Rightarrow 4$. The other path is the reverse, from the $HOMO^{\downarrow}$ of $Co4$ to the $LUMO^{\downarrow}$ of $Co2$, namely $4\Rightarrow 2$. Along the $c$-axis there are two more $HOMO^{\downarrow}$-$LUMO^{\downarrow}$ paths, represented by $2\mapsto 2$ and $4\mapsto 4$.  In $Fe_3O_2BO_3$, there is just $2\Rightarrow 4$ in the triad and $2\mapsto 2$ along $c$, as there are no spin down electrons in the $t_{2g}$ levels of the $Fe4$ monomer. Table 4 shows the inter-site transition energies $\Delta_{ij}$=$E_{LUMO^{\downarrow}_j}$-$E_{HOMO^{\downarrow}_i}$, from which energetic trends could be obtained.

\vskip.5in

\begin{tabular}{|c|c|c|c|c|c|}
\hline
\multicolumn{3}{|c|}{{\bf Table 4} - Direct intersite $HOMO_i$-$LUMO_j$  }\\
\multicolumn{3}{|c|}{transition energies $\Delta_{ij}$ in the triad. Energies in eV.} \\ 
\multicolumn{3}{|c|}{"U" parameter, discussed in the text.}\\ \hline
site pair & $Co_3O_2BO_3$ & $Fe_3O_2BO_3$ \\
$i, j$&   &  \\ \hline
$2\Rightarrow 4$ & $0.16$ &  $0.05$ \\ \hline
$4\Rightarrow 2$ & $-0.06$ + $"2U"$& - \\ \hline
$2\mapsto 2$ & $0.01$ + $"U"$& $0.04$ + $"U"$   \\ \hline
$4\mapsto 4$ & $0.06$ + $"U"$& - \\ \hline
\end{tabular}
\vskip.5in

Effects of in-site electron-electron repulsion on inter-site transfer is qualitatively taken into account in Table 4 by hypothetical amounts $"U"$ and $"2U"$, aiming to indicate differences which would arise upon changing the oxidation state of the pair. For instance, the $2\Rightarrow 4$ jump changes the atoms oxidation states from $2^+, 3^+$ to $3^+, 2^+$, with no net effect, so that no repulsive amounts were added to the electron-lattice contribution to this jump ($0.16eV$ for $Co$ and $0.05eV$ for the $Fe$ ludwigite). On the other hand, the $4\Rightarrow 2$ jump changes the pair oxidation states from $3^+,2^+$ to $4^+,1^+$, so that repulsion energy increases are represented by $"2U"$. The $2\mapsto 2$ jump changes the pair oxidation state from $2^+$, $2^+$ to $3^+$,$1^+$, thus an amount of $-"U"+"2U"$=$"U"$ has been added. The analysis would suggest that electron hopping in $Co_3O_2BO_3$ deals with higher energetic cost than in $Fe_3O_2BO_3$.


It is possible to gain additional insight by investigating the electronic structure of molecular dimers, since dimerization is the essential physical mechanism of the structural transition in $Fe_3O_2BO_3$. As metal-metal interaction is directly influenced by oxygen ligands, one considers the subunits $Co_2Co_4O_{10}$ and $Fe_2Fe_4O_{10}$, which preserve the octahedral vicinity of the metal pair. In Fig.4, it is shown the calculated molecular orbital structure of these metal dimers, seen as built up from individual fragments $MO_5$; three $t_{2g}$ levels of each fragment, $M2O_5$ and $M4O_5$, combine to form the six lower lying molecular orbital levels. As the constituent fragments have electronic configurations corresponding to $Co^{2+/3+}$ and $Fe^{2+/3+}$, the three lowest (for $Co$) and one lowest (for $Fe$) dimer molecular orbitals (MO's) are occupied with spin down electrons. 

One would expect more covalent dimers to be more likely to dimerize in the crystalline structure. Since energy levels of neighbor $Fe$ monomers are closer to each other, when compared to $Co$ monomers (see Fig.3), better conditions for dimerization are to be expected in $Fe_3O_2BO_3$. This is in fact confirmed by actual calculation. As the lowest dimer MO's are ligand $3d$-$O_{2p}$ combinations, covalency could be estimated by means of minority spin density distribution in the subunits. Let $\rho_{E_i}(atom)$ be the atom-projected density of states in MO energy level $E_i$. The lower lying MO's $i=1,2,3$ in $Co$ and $i=1$ in $Fe$ subunits, contribute with $\rho_{E_i}(2)$ and $\rho_{E_i}(4)$, so that the ratio $r_{2,4}$=$[\sum_i{\rho_{E_i}(2)}]/[\sum_i{\rho_{E_i}(4)}]$ provides the degree of localization of minority spins within the dimer. Calculations give a density ratio of $5.6$ for cobalt, significantly larger than the corresponding ratio found for $Fe$ dimers; in the latter, $r_{2,4}$=$1.3$. Thus, in the dimer $Co2Co4O_{10}$, minority spin density is mainly localized around $Co2$, while in the $Fe2$-$Fe4$ dimers the extra spin spreads almost equaly among the two metals, a situation which favors dimerization, as compared with the $Co$ compound. 

Outside the triad, monomers $1$ and $3$ in $Co_3O_2BO_3$ have molecular electronic structures fairly equal to that of site $2$, with the $t_{2g}$ levels spreading between $-13 eV$ and $-12.8 eV$ below the corresponding levels of $Co4$. Thus, all $Co^{2+}$ sites are structurally analogous, and distinct from $Co^{3+}$. In $Fe_3O_2BO_3$, metal sites $1$ and $3$ have $t_{2g}$ levels spreading in the same energy range as that of the triad sites, but the $2+$ configuration is slightly more stable.

\section{Discussion}

Charge distribution in $Co$ and $Fe$ homometallic ludwigites show localization in the former and delocalization in the latter. This difference is related to local distortions of the four oxygen octahedra which seem to determine specific conditions for conductivity paths. Electronic equivalence of $Fe$ sites in the $Fe_3O_2BO_3$ triad is an important aspect of the difference between the electronic structure of the two materials.

Electronic structure equivalence of $Fe^{2+}$ and $Fe^{3+}$ monomers have also been found in the mixed valent $Fe_2OBO_3$ warwickite\cite{mm-warwfe}. Taking into account that both $Fe$ oxyborates have structural transitions associated to charge rearrangements - through dimerization in the ludwigite\cite{larrea-local} and by means of a Wigner phase in the warwickite\cite{nature}, one could establish a common behavior in these two materials. In both cases, easy paths for electron transfer between di- and tri-valent sites create instabilities in the charge distribution, which allows different arrangements. If other stabilizing mechanisms are present, charge ordering/localization may occur in such conditions. In the $Fe$ warwickite this is provided by coulombian repulsion between the extra electrons of $Fe^{2+}$, defining the Wigner crystal arrangement of the low temperature phase\cite{nature}. In the ludwigite, dimerization provides chemical stabilization. In a recent work, Vallejo et al.\cite{vallejo} have shown that local ferromagnetic ordering in the triad could strenghten $Fe2$-$Fe4$ bonds, contributing further to stabilize dimerization. It should be noted that in the other two mixed-valent oxyborates, $Co_3O_2BO_3$ and the warwickite $Mn_2OBO_3$, no structural transition has been observed. A similar theoretical analysis has found larger barriers for electron jump in the manganese compound, associated to Janh-Teller distortions\cite{mm-warwmn}.  

Larger barriers for electron jump in $Co_3O_2BO_3$, found in the analysis above, would be consistent with a Mott like conductivity regime, as suggested by Ivanova et al.\cite{ludwco2}. The band structure calculation of Fig.3 predicts, on the other hand, metallic behavior (see Fig.3), as suggested by Freitas et al.\cite{ludwco1}. More experimental and theoretical studies are necessary to determine the electrical conductivity regime of $Co_3O_2BO_3$. The present investigation could provide some basic understanding of the role of electron-lattice interaction in the compound.  

\section{Conclusion}

In this paper, a basic structure-property investigation is carried out in two homometallic ludwigites, $Co_3O_2BO_3$ and  $Fe_3O_2BO_3$, by using the EHTB-{\it hsf} method. The band structures of $Co_3O_2BO_3$ and of two crystalline phases of $Fe_3O_2BO_3$ were found to be very similar, consistently with the structural similarity of the compounds. Different $3d$ occupancy of $Co$ and $Fe$ leads however to important differences in the electronic structures of both materials. In both compounds short metal-metal distances in the triad generate a separation of the lower $t_{2g}$ band. However, due to electron occupancy, it is only in $Fe_3O_2BO_3$ that the separation constitutes a true 3LL semiconductor gap, consistently with experiment. 

Calculations show that small differences in the local octahedral geometry at different sites lead to significant differences in the charge distribution of the two compounds. In agreement with M\"ossbauer results, electronic equivalence was observed between divalent and trivalent $Fe$ cations in the triad, at both structural phases of $Fe_3O_2BO_3$. On the other hand, structural and electronic conditions for charge localization are obtained in $Co_3O_2BO_3$.  The electronic structure of all relevant monomers and dimers were examined, showing that site non equivalence could create higher barriers for electron hopping in the cobalt ludwigite, hindering dimerization in this compound. For $Fe_3O_2BO_3$, charge distribution instabilities, related to small hopping barriers, are suggested to play a relevant role in the structural transition observed for this material.

An analogy is made with the two mixed-valent warwickites, $Fe_2OBO_3$ and $Mn_2OBO_3$, since site equivalence in the former has been associated to the structural transition and changes in the conductivity regime, while, in the $Mn$ compound, structural stability is associated with structural charge ordering. The nature of electrical conductivity in $Co_3O_2BO_3$ is briefly discussed. 

For a complete understanding of the physics of the mixed-valent ludwigites, more experimental and theoretical work is necessary. The present results could provide a comprehensive understanding of important features related to electron-lattice interactions in these complex materials.

\end{document}